\documentclass[a4paper,fleqn]{cas-dc}

\usepackage[numbers]{natbib}
\usepackage{adjustbox}
\usepackage{gensymb}
\usepackage{tabu}
\usepackage{tabularx}
\usepackage{graphicx}
\usepackage{dcolumn}
\usepackage{natmove}
\usepackage{microtype}
\def\tsc#1{\csdef{#1}{\textsc{\lowercase{#1}}\xspace}}
\tsc{WGM}
\tsc{QE}
\tsc{EP}
\tsc{PMS}
\tsc{BEC}
\tsc{DE}
\begin{document}
\shorttitle{Crystal Structure of $\tau_{11}$ phase}
\shortauthors{Biswas Rijal et~al.}

\title [mode = title]{Crystal Structure of the $\tau_{11}$ Al$_4$Fe$_{1.7}$Si Phase from Neutron Diffraction and Ab Initio Calculations}                      
%%%\tnotemark[1,2]

%%%\tnotetext[1]{This document is the results of the research
 %%%  project funded by the Department of Defense.}

%%%\tnotetext[2]{The second title footnote which is a longer text matter
 %%%  to fill through the whole text width and overflow into
 %%%  another line in the footnotes area of the first page.}

\author{Biswas Rijal$^1$}

\author{Sujeily Soto$^1$}

\author{Kausturi Parui$^1$}
\author{Anil Sachdev$^2$}
\author{Megan M. Butala$^1$}
\author{Michele V. Manuel$^1$}
\author{Richard G. Hennig$^{1,}$}\corref{cor1}

\address{$^1$ Material Science and Engineering, University of Florida, Gainesville, FL 32611}
\address{$^2$ General Motors Research \& Development, Warren, MI}

\cortext[cor1]{Electronic mail: rhennig@ufl.edu}

\begin{abstract}
%The aim of the present study was to determine the crystal structure of Al$_4$Fe$_{1.7}$Si ($\tau_{11}$) phase. The crystal structure of $\tau_{11}$  phase was derived by combination of powder neutron diffraction and density functional theory calculations. The neutron diffraction data was refined using Pawley and Reitveld refinement to obtain the crystal structure information. The composition of the $\tau_{11}$ phase was Al 64.7 Fe 25.1 Si 10.3 . $\tau_{11}$ phase has hexagonal crystal structure with the lattice parameters a=7.478 and c=7.472.
The intermetallic $\tau_{11}$ Al$_4$Fe$_{1.7}$Si phase is of interest for high-temperature structural application due to its combination of low density and high strength. We determine the crystal structure of the $\tau_{11}$ phase through a combination of powder neutron diffraction and density functional theory calculations. Using Pawley and Rietveld refinements of the neutron diffraction data provides an initial crystal structure model. Since Al and Si have nearly identical neutron scattering lengths, we use density-functional calculations to determine their preferred site occupations. The $\tau_{11}$ phase exhibits a hexagonal crystal structure with space group $P6_3/mmc$ and lattice parameters of $a=7.478$~\AA\ and $c=7.472$~\AA. The structure comprises five Wyckoff positions; Al occupies the $6h$ and $12k$ sites, Fe the $2a$ and $6h$ sites, and Si the $2a$ sites. We observe site disorder and partial occupancies on all sites with a large fraction of 80\% Fe vacancies on the $2d$ sites, indicating an entropic stabilization of the $\tau_{11}$ phase at high temperature.
\end{abstract}


\maketitle

\section{Introduction}

Al-Fe based compounds possess high melting points, high hardness, low density, low cost, and good oxidation and corrosion resistance, making them highly attractive structural materials. A high Al content in Al-Fe alloys is desirable for lightweight engine parts to drive high power densities and excellent transient engine performance~\cite{liu2012ordered}. However, the mechanical properties of the Al-Fe compounds tend to decrease with increasing Al content. The application of high Al-content Al-Fe intermetallic compounds is especially limited by their brittleness. For example, the compressive strain drops rapidly from 0.8\% at 560 MPa (Fe$_3$Al, face-centered cubic structure) to 0\% at 200 MPa (FeAl$_3$, monoclinic structure)~\cite{liu2012ordered}. 
 
However, the addition of Si into an Al-Fe binary system stabilizes a crystal structure with low density and improved mechanical properties. The Al-Fe-Si ternary phase space is complex, consisting of at least 11 equilibrium ternary phases and 19 invariant reactions~\cite{du2008thermodynamic}. In addition, at least 5 metastable ternary phases have been reported~\cite{ghosh2008aluminium}. Al$_4$Fe$_{1.7}$Si, known as $\tau_{11}$, holds great promise for structural applications as a result of its low theoretical density, about 4.1~g/cm$^3$, unique hexagonal crystal structure and an estimated tensile strength of 1500 MPa.~\cite{kim2013removal,liu2019rapidly}

\begin{figure*}
    \centering
    \includegraphics[width=\textwidth]{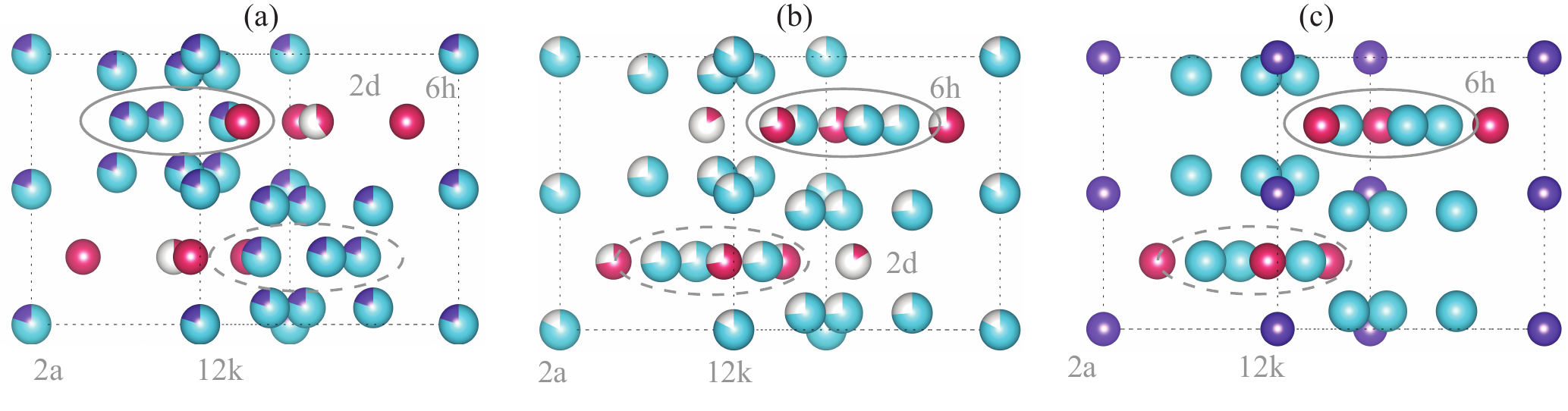}

\hspace{-0.4cm}
\begin{tabular}{m{.5cm}m{.5cm}m{.72cm}m{.62cm}m{.62cm}m{.62cm}}
Wyck. & At. & Occ. & $x$ & $y$ & $z$  \\
\midrule
2$a$ & Al/Si & 0.8/0.2& 0 & 0 & 0 \\
6$h$ & Al/Si & 0.8/0.2 &  0.4620 & 0.9240 & 0.25 \\
12$k$ & Al/Si & 0.8/0.2 &  0.2 & 0.4 & 0.4385\\
6$h$ & Fe & 1.0 & 0.1209 & 0.2418 & 0.75 \\
2$d$ & Fe & 0.4 & 0.3333 & 0.6667 & 0.75\\
 & a~=~7.509~\AA & & c~=~7.594~\AA
\end{tabular}
\hspace{0.1cm}
\begin{tabular}{m{.5cm}m{.3cm}m{.62cm}m{.62cm}m{.62cm}m{.4cm}}
Wyck. & At.& Occ. & $x$ & $y$ & $z$  \\
\midrule
2$a$ & Al & 1.0 &  0 & 0 & 0\\
6$h$ & Al& 0.90 &  0.5393 & 0.0785 & 0.25 \\
12$k$ & Al& 0.90 &  0.1997 & 0.3994 & 0.4380 \\
6$h$ & Fe & 0.87 &  0.1216 & 0.2433 & 0.75 \\
2$d$ & Fe & 0.20 &  0.33 & 0.66 & 0.75\\
& a~=~7.518~\AA & & c~=~7.571~\AA
\end{tabular}
\hspace{1mm}
\begin{tabular}{m{.5cm}m{.3cm}m{.62cm}m{.62cm}m{.62cm}m{.2cm}}
Wyck. & At. & Occ. & $x$ & $y$ & $z$\\
\midrule
2$a$ &Si& 1.0 &  0 & 0 & 0 \\
6$h$ &Al & 1.0 &  0.5417 & 0.0834 & 0.25\\
12$k$ &Al & 1.0 &  0.1994 & 0.3987 & 0.4349\\
6$h$ &Fe & 1.0 &  0.1223 & 0.2446 & 0.75\\
 & & & & &\\
 & a~=~7.479~\AA & & c~=~7.473~\AA
\end{tabular}
\caption{\label{fig:structure} (Color online) The crystal structure of the $\tau_{11}$ phase from (a) Ref.~\cite{german1989crystal}, (b) the neutron diffraction refinement, and (c) the DFT relaxations. The aqua, purple, and red spheres denote Al, Si, and Fe, respectively. Two colors in a sphere indicates mixed occupancy of two atomic species, and white partial occupancy depicts partial occupation of a site. Partial occupancy used in DFT calculations in the different sites were motivated by experimentally observed probabilities. The Wyckoff positions are labeled, and the ellipses indicate the shift in the Al2 ($6h$) sites along the $y$ coordinate between the previous refinement and this work. Three tables below denote the atomic positions and occupancies of the three models, i.e., Ref.~\cite{german1989crystal}, the Rietveld refinement, and DFT relaxations.}
    
\end{figure*}

For the Al-Fe-Si system, x-ray and neutron diffraction cannot easily distinguish between Al and Si due to their similar x-ray form factor and neutron scattering length. Therefore, we combine a refinement of experimental
neutron diffraction data with density functional theory (DFT)
calculations. Figure~\ref{fig:structure}(a) illustrates that the previously reported crystal structure of the $\tau_{11}$ intermetallic phase based on x-ray diffraction~\cite{german1989crystal} assumed a mixed occupancy of Al and Si with the ratio given by the overall composition and partial occupancy of Fe. The $2a$, $6h$, and $12k$ Wyckoff sites have 80\% Al and 20\% Si occupancies and the $2d$ site has a partial occupancy of 45\% Fe. Computational methods provide an opportunity to determine the energy of different site occupancies.

In this work to better understand and model the properties of the $\tau_{11}$ phase, we combine a refinement of experimental neutron diffraction data with density functional theory (DFT) calculations to investigate the structure and site occupancies. The results reveal a different atomic structure than that previously reported and identify preferred Al occupancy on the $6h$ and $12k$ sites and Si occupancy on the $2a$ site. In our experimental approach, while the lattice constants and symmetry of the published structure reproduced the positions of peaks in diffraction data, considerable changes in atomic positions were required to model peak intensities.
%We could not differentiate site occupancy by Al and Si experimentally, owing to their nearly identical scattering length densities, however, we do report a more accurate structural model for $\tau_{11}$.
DFT calculations confirm that this structure has an energy 250~meV/atom lower than the previously published model. Noteworthy, we obtained the experimental and computational structural models independently, initializing the refinement and structural relaxation in each approach with the atomic positions of the published structure (Figure\,\ref{fig:structure}(b) and (c)). In addition to consistency from parallel approaches, the resulting structure is more intuitive, with a more homogeneous distribution of atoms across the structure. This improved atomic structure model enables future alloy design studies to improve the high-temperature thermodynamics stability, identify optimal synthesis conditions, and enhance the mechanical properties of $\tau_{11}$.

\section{Methods}

\subsection{Synthesis}
Al-Fe-Si ternary alloys with nominal compositions of Al$_{64.0}$Fe$_{25.0}$Si$_{11.0}$ and Al$_{63.0}$Fe$_{25.5}$Si$_{11.5}$ were fabricated by arc melting 99.99 wt.\% Al and 99.98 wt.\% Fe from Sigma Aldrich with an Al-50 wt.\% Si alloy from Belmont Metals under an argon atmosphere. The melting was repeated 5 times to ensure homogenization of the alloys. Then the alloys were annealed at 950$^{\circ}$C for 100\,h. In order to protect the samples from oxidation at high temperatures, each alloy was individually wrapped inside VakPak65 heat treating containers. After the heat treatments, the containers with the samples inside were quickly removed from the furnace and quenched in water. A Tescan MIRA3 scanning electron microscope (SEM) coupled with an EDAX Octane Pro energy dispersive spectrometer (EDS) was used to measure the chemical composition of the alloys. The average of at least three points were measured to obtain the compositions. 

\subsection{Neutron Diffraction and Analysis}
Alloy samples were ground into fine powders for neutron diffraction experiments. The neutron powder diffraction (NPD) was measured at HFIR-HB2A at Oak Ridge National Laboratory. The measurements were performed at room temperature with a constant wavelength of 1.54 Å over an 8$\degree$\,<\,2$\theta$\,<\,154$\degree$ range.

For the NPD data analysis, GSAS-II~\cite{toby2013gsas} and Topas Academic v6~\cite{coelho2018topas} software were used for initial LeBail and Pawley fits. Lattice and instrumental parameters from these fits were used in the subsequent Rietveld refinement of the full unit cell details of lattice parameters, atomic positions, atomic displacement parameters, and site occupancies. Since the scattering lengths densities of Al of 2.078$\times$10$^{-6}$\AA $^{-2}$ and Si of 2.074$\times$10$^{-6}$\AA$^{-2}$ are nearly identical, we used a single species on sites that realistically have mixed occupancy of Al and Si. The resulting error in the structure factor from this simplification is minor and reduces the number of free parameters to give a more meaningful fit.

\subsection{Computation}
We used DFT energies of various configurations to distinguish the site occupation for Si and Al, which is not resolved by diffraction alone. The computational prediction of the ground state structure for a given composition involves finding the structure or set of structures with the lowest formation energy at that composition. We construct the convex hull of the energies for the known and calculated Al-Fe-Si phases to compare the energy of structures with different compositions and identify the most stable site occupations. The convex hull is defined as the set of points that encloses all the points in the set. The enclosing points can be used to determine the lowest energy structure.  For the convex hull, lines and planes connect the lowest energy phases and represent the system's energy at 0\,K.
%Structure computation requires the minimization of free energy from the construction and minimization of a convex hull curve.
%Once we construct the convex hull diagram of the ground state energies of various structures and phases, the
Configurations with energies above the convex hull are unstable. The distance from the hull for various structures indicates the relative stability of those structures.

The structural relaxation and energy calculations were performed with the plane-wave DFT code VASP~\cite{kresse1993ab, kresse1994ab, kresse1996efficient, kresse1996efficiency} using the PBE functional~\cite{perdew1996generalized} and the projector augmented wave method~\cite{blochl1994projector, kresse1999ultrasoft}. The cutoff energy for the plane wave basis set was set to 450\,eV with a $k$ point density of 1,000 points per reciprocal atom. We used very tight convergence criteria for the structure relaxations. All structures were relaxed until the energy difference between subsequent electronic steps was less than 0.001 meV and between the ionic relaxation steps the difference was less than 0.01 meV. 

\section{Results and Discussion}
\subsection{Energy minimization}

To create the convex hull, we obtain the crystal structures of all potentially competing phases for $\tau_{11}$-Al$_4$Fe$_{1.7}$Si from the Inorganic Crystal Structure Database (ICSD)\cite{levin2018nist} and the 
Materials Project database~\cite{Ong2008, Jain2011a, Jain2013}. We do not consider the disorder in the competing phases and used the crystal structures as provided by the Materials Project database. Then, we relax all the structures and calculate their energies. The considered Al-Fe-Si phases include the elemental phases of Al, Fe, and Si, and 12 binary and 36 ternary compounds.
The crystal structure of $\tau_{11}$-Al$_4$Fe$_{1.7}$Si (initially called $\tau_{10}$) from German \textit{et al.}~\cite{german1989crystal} exhibits a hexagonal unit cell of the Co$_2$Al$_5$-type structure with 28 atoms, space group $P6_3/mmc$, and lattice parameters $a = 7.509$~\AA~ and $c = 7.594$~\AA. The 2$a$, 6$h$, and 12$k$ Wyckoff sites are partially occupied by a mixture of Al and Si, and Fe occupies the 6$h$ and, partially, the 2$d$ sites.

\begin{figure*}
    \center
    \includegraphics[width=1.0\textwidth]{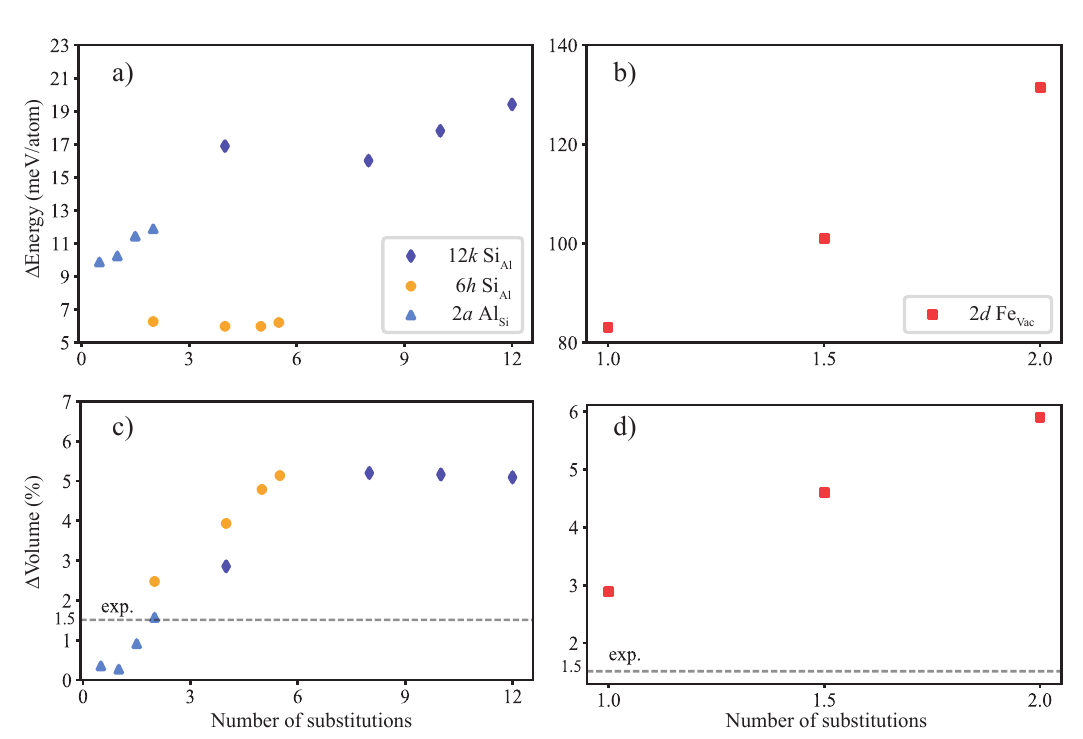}
    \caption{\label{fig:hull_distance} [a,b] Change in energy for various site substitutions relative to the $\tau_{11}$-Al$_4$Fe$_{1.7}$Si structure with the lowest energy. The energy, $E_\mathrm{hull}$ in meV/atom is measured by the distance from the convex hull obtained from all possible competing phases of the $\tau_{11}$ phase. [c,d] Percentage change in volume of the various choices of site occupancy for the $\tau_{11}$-Al$_4$Fe$_{1.7}$Si structure and the lowest energy structure. The dashed line represents the experimental volume obtained from the refinement. The symbols in all the figure represent the substitution in a Wyckoff substitution.}
\end {figure*}

%We{\it The structure from German \textit{et al.} reported partial and mixed site occupancies for Al and Si; in order to respect the atomic ratio in our computational structures, we added 2 Si atoms in 2$a$ Wyckoff positions.}{\bf [RGH: The previous sentence seems out of place here and I removed it.}
Figure~\ref{fig:hull_distance} a,b shows the change in hull distance between the lowest energy structure and the structures with site substitution in 2$a$, 6$h$, 12$k$, and 2$d$ Wyckoff sites by Al, Fe, and Si. The energies are obtained from DFT energies relative to the convex hull of competing phases. In the DFT convex hull, the competing phases for the $\tau_{11}$ phase are Al$_{26}$Fe$_9$Si$_6$, Al$_{13}$Fe$_4$ and Al$_2$Fe$_3$Si$_3$.
%We relaxed the atomic positions and lattice vectors to obtain the lowest energy structure.

The DFT calculations provide the insight into occupational disorder on the Wyckoff sites. The refinement cannot resolve the occupation of the 2$a$, 12$k$, and 6$h$ Wyckoff sites by Al and Si and indicates a small partial occupancy of the 2$d$ site by Fe. The hull distance is smallest when Si occupies the 2$a$ sites, Al the 12$k$ and 6$h$ sites, and the 2$d$ site is empty. The energy cost for Al occupying a 2$a$ site is about 10\,meV/atom. The energy cost for a Si atom replacing Al on one of the 6$h$ sites is about 10\,meV/atom, and for the 12$k$ site, it is slightly higher at about 20\,meV/atom. This indicates that Si preferentially occupies the 2$a$ and Al the 6$h$ and 12$k$ Wyckoff sites. 
For the 2$d$ site, occupying it by Fe increases the energy by about 100\,meV/atom, indicating that the 2$d$ site will be preferentially empty even at high temperatures, consistent with the diffraction analysis by German \textit{et al.}~\cite{german1989crystal}.
The small energy cost for Al/Si occupational disorder indicates that configurational disorder stabilizes the $\tau_{11}$ phase at high temperatures and that Si preferentially occupies the 2$a$, Al the 6$h$ and 12$k$, and Fe the other 6$h$ site, while the 2$d$ site is preferentially empty. 

In addition to the energy, we also compare the Wyckoff positions and lattice parameters of the different configurations with the experimental ones. Most Wyckoff positions are very similar between the DFT relaxed structure and the structure by German \textit{et al.}. However, we find a large shift of the Al atoms in the 6$h$ position by close to 1\AA\ along the $y$-axis, which is illustrated in Fig.~\ref{fig:structure}. As shown below, this change in position is consistent with our neutron diffraction data. For the lattice parameter, German \textit{et al.}~\cite{german1989crystal} obtained $a = 7.509$~\AA\ and $c = 7.594$~\AA. For the lowest energy structure shown in Figure~\ref{fig:structure}(c), we obtain similar values of $a = 7.478$~\AA\ and $c = 7.472$~\AA. Figure~\ref{fig:hull_distance}c,d shows the change in volume of the unit cell due to site occupancies. The experimental volume is 1.5\% larger than the DFT calculated volume for the lowest energy structure. We observe that the volume shows larger changes when Si substitutes Al in the 6$h$ and 12$k$ positions. Likewise, the change in volume is between 3-5\% when Fe occupies the vacancy in the 2$d$ position. Since the PBE functional typically overestimates the volume, the increase in volume by about 3 to 5\% percent due to Al/Si disorder on the 6$h$ and 12$k$ sites and Fe/vacancy disorder on the 2$d$ sites is consistent with the DFT results.

%{Our calculated structure also has a $c$/$a$ ratio of 0.99 and a density of 4.0222~g/cm$^3$, which are comparable to the pure Ti metal, with a $c$/$a$ ratio of 1.588 and a density of 4.55\,g/cm$^3$~\cite{oh1989relationship}. The packing fraction of the predicted structure is 0.77 indicating it has a very closed packed structure.
%Primary differences from the published structure are in the positions of Al/Si atoms in $y$ for the 6$h$ Wyckoff positions (Figure\,\ref{fig:structure}a and b) and the absence of Fe atoms in 2$d$ Wyckoff sites.

Thus, the DFT calculations show that the $\tau_{11}$ phase has the lowest energy with an ordering of Al and Si such that Al occupies the 12$k$ and 6$h$ positions and Si the 2$a$ position, and Fe occupying another 6$h$ site.  In addition, our calculations indicate that a vacant 2$d$ position is energetically favorable. Therefore, DFT predicts the lowest energy $\tau_{11}$ structure with 18 Al, 6 Fe, and 2 Si atoms in 6$h$ and 12$k$, 2$d$, and 2$a$ positions, respectively. The predicted lattice parameters match the published structure. However, there is a significant shift of 1~\AA\ in the 6$h$ position compared to the published structure. Therefore, we conducted a neutron diffraction study of the $\tau_{11}$ structure to validate the DFT results.
In this study, we were concerned with determining the lowest energy crystal structure and neglect the role of entropy on the disorder at higher temperatures.

\subsection{Structure Refinement}

\begin{figure}
    \center
    \includegraphics[width=1.0\columnwidth]{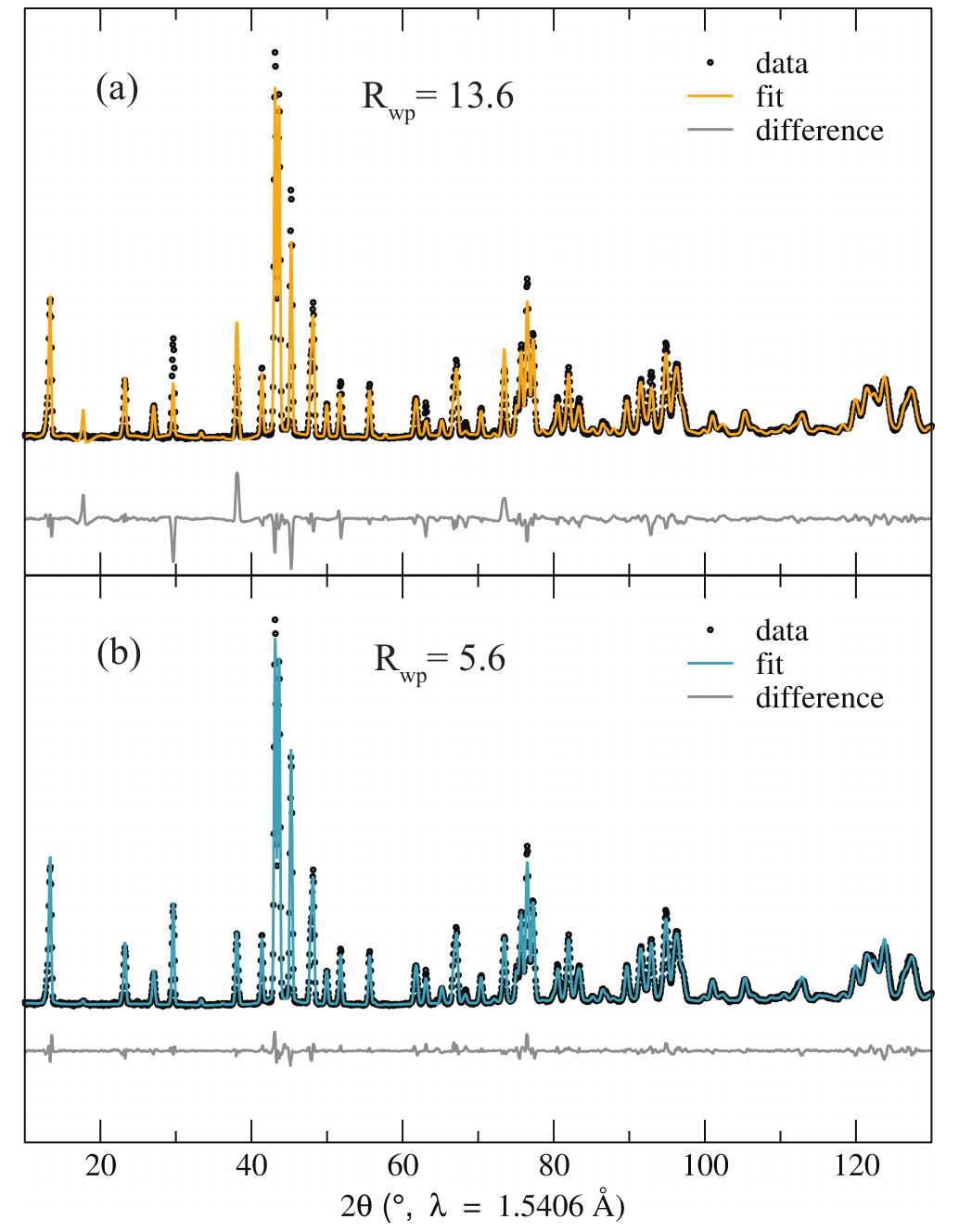}
    \caption{\label{fig:Refinement} Comparative plots of data, fit and difference curve with Fe on the original (a) vs. refined site (b). Neutron diffraction data (dots) are fitted (orange and blue) using Rietvield refinement. The difference (black) line shows the difference between the actual data and the fit. Correcting the Fe site significantly improves the goodness of from $R_{wp}=13.6$ to $R_{wp}=5.6$.} 
\end {figure}

Given the considerable difference between the lowest energy structure we identified with DFT and that previously published, we refined the atomic structure model (beginning with the published structure) against experimental data obtained from NPD. EDS measurements at various points gave a mean composition of 64.7 at.\% Al, 25.1 at.\% Fe and 10.3 at.\% Si (and the relevant error/standard deviation) as shown in Table~\ref{tab:table2}. DFT calculated the lowest energy structure for the $\tau_{11}$ phase with  69.23 at.\% Al, 23.07 at.\% Fe 7.69 at.\% Si. The NPD pattern shown in Figure~\ref{fig:Refinement} has the peak positions that could be matched to the expected reflections for the published $\tau_{11}$ phase, confirming  the single phase microstructure observed in the SEM image. 

\begin{table}
\caption{\label{tab:table2}Composition obtained from EDS micrograph in atomic percent with standard deviation inside the parenthesis.}
\centering
\begin{tabular}{m{.5cm}m{.5cm}m{.5cm}|m{.5cm}m{.5cm}m{.5cm}}
Nominal Phase & & & Measured Phase & &\\
composition & & & composition  & &\\
\midrule
Al & Fe & Si & Al & Fe & Si\\
64 & 25 & 11 & 64.7 & 25.1 & 10.3\\
%& & & (0.02) & (0.04) & (0.04)\\
% & & & & &
\end{tabular}
\end{table}

During initial Rietveld refinement (using GSAS-II), we found a considerable mismatch between the peak intensities of the published structure and our NPD data. With the various parameters in the unit cell, particularly the mixed and partial site occupancies and degrees of freedom on several atomic position, we were initially in a local least squares minima, preventing us from capturing the relevant structural details. Throughout this effort, peak mismatch, and a high $R_{wp}$> 20 persisted, indicating a substantial difference between our initial model (based on the published structure) and that of our sample. 

This initial pass indicated that while the symmetry and lattice parameters, which give rise to the peak positions, were correct, we would need to refine the unit cell model to capture peak intensities. To confirm our lattice parameters and determine instrumental parameters, we performed a Pawley refinement (using Topas). Pawley refinement is a structure-free approach that generates d-spacings based on the lattice parameters and Miller indices, and modifies these to match peak position. However, with Pawley fitting, the intensity of each peak comes from a unique parameter, with no tie to a structural parameter. 

In contrast, in the Rietveld refinement, the complete unit cell (lattice parameters, occupancies, atomic displacement parameters, and, particularly, atomic positions) is refined and the positions and identities of atoms in the structural model are used to calculate peak intensities in the fit profile. Beginning with the lattice and instrument parameters (e.g., low angle peak asymmetry correction, zero-offset, etc.) from Pawley refinement and the atomic positions and occupancies from the published structure, we again saw significant peak intensity mismatch in Rietveld fitting, as expected. 

Before refining atomic positions, occupancies, and atomic displacement parameters ($B_{eq}$), we modified the unit cell to reduce the number of correlated parameters. In particular, we simplified our model based on the mixed Al and Si occupancy on 2$a$, 6$h$, and 12$k$ sites~\cite{german1989crystal}; in particular, given their nearly identical scattering lengths (2.078$\times$10$^{-6}$\AA $^{-2}$ for Al and 2.074$\times$10$^{-6}$\AA$^{-2}$ for Si), these species cannot be distinguished using typical diffraction methods. As such, there was no physical insight to draw from refining their relative occupancies in the structure. Accordingly, we modeled the mixed Al—Si sites as occupied by only Al (Si alone would produce functionally equivalent results). The resulting error in the structure factor from this simplification was be minor and resulted in a more meaningful fit with fewer correlated parameters.

The fit improved upon refining the lattice parameters, atomic positions, occupancies, and isotropic displacement parameters ($B_{eq}$), yielding an $R_{wp}$ of 13.6 (Figure\,\ref{fig:Refinement}a). The most considerable difference between the model and the experimental data was in the $y$ positions of Al2 (on a 12$k$ site), which is in agreement with the atomic position identified by DFT. In the interest of, again, a more homogeneous distribution of atoms in the unit cell, we shifted Fe1 to an alternate 2$d$ site, with $z$\,=\,$\frac{3}{4}$ rather than $\frac{1}{4}$. In combination with the refinement of a Gaussian strain parameter, which improved the peak shape fit at high angles, this change in Fe1 position resulted in excellent agreement between the data and the structural model, reflected in an $R_{wp}$ of 5.2 (Figure\,\ref{fig:Refinement}b). The final positions and occupancies from this fit are detailed in Figure\,\ref{fig:structure} and, aside from a second and low occupancy Fe site, shows excellent agreement with the lowest energy structure identified from DFT.

To ensure the physical relevance of our refined model, particularly the resulting atomic ratios, we used the occupancies and site multiplicities to compare the ratio of main group elements (Al+Si) and Fe, since the ratios of Al and Si could not be individually refined in the model. The resulting (Al+Si)/Fe ratios from the refinement and the experimentally measured composition are 3.2 and 3.0 respectively. This confirms the physical relevance of the refined structure.

Our calculated and refined structures have a $c$/$a$ ratio of 0.99 and a density of 4.02~g/cm$^3$, which are lower than those of pure Ti metal, with a $c$/$a$ ratio of 1.588 and a density of 4.55\,g/cm$^3$~\cite{oh1989relationship}. The packing fraction of the predicted structure is 0.77 indicating it has a very closed packed structure. Primary differences from the published structure are in the positions of Al/Si atoms in $y$ for the 6$h$ Wyckoff positions (Figure\,\ref{fig:structure}a and b) and the absence of Fe atoms in 2$d$ Wyckoff sites.

\section{Conclusion}
Through the combination of DFT calculations and Rietveld refinement of neutron diffraction data, we determined a revised structure of the  $\tau_{11}$-Al$_4$Fe$_{1.7}$Si intermetallic phase. The most significant change is the shift in the positions of the Al/Si 6$h$ position, which was consistently identified in the neutron diffraction refinement and the DFT calculations, reducing the energy by 260\,meV compared to the previously reported structure~\cite{german1989crystal}. Since the neutron scattering length of Al and Si are nearly indistinguishable, we used DFT calculations to identify preferential site occupations. Based on the DFT energies, we predict that Al and Si are disordered and that Al prefers the 6$h$ and 12$k$ sites and Si the 2$a$ site.  The observation of a large fraction of Fe vacancies and the Al/Si disorder indicates an entropic stabilization of the $\tau_{11}$-Al$_4$Fe$_{1.7}$Si phase at high temperature.

From a chemical perspective, the new structure is more intuitive, with a more homogeneous distribution of density across the unit cell. This improved structural model retains its $c/a$ ratio with highly closed packed structure. Further, variations of the structure and computational prediction of their properties are enabled, allowing improved determination of structure-property relationships and property prediction. Studies can be conducted to increase the phase stability region for $\tau_{11}$ with quaternary addition to enable alloys that are still low cost, low density, and suitable for high temperature applications, replacing more costly Ti-based alloys. Our results additionally highlight the utility of modern diffraction and structural modeling algorithms to advance structure-property understanding of metal alloys.

\section {Disclaimer}

This report was prepared as an account of work sponsored by an agency of the United States Government.  Neither the United States Government nor any agency thereof, nor any of its employees, makes any warranty, express or implied, or assumes any legal liability or responsibility for the accuracy, completeness, or usefulness of any information, apparatus, product, or process disclosed, or represents that its use would not infringe privately owned rights.  Reference herein to any specific commercial product, process, or service by trade name, trademark, manufacturer, or otherwise does not necessarily constitute or imply its endorsement, recommendation, or favoring by the United States Government or any agency thereof.  The views and opinions of authors expressed herein do not necessarily state or reflect those of the United States Government or any agency thereof. 

\section{Acknowledgement}
This materials is based upon work supported by the U.S. Department of Energy (DOE), Office of Energy Efficiency and Renewable Energy (EERE), specifically the Vehicle Technologies Office under Award Number DE-EE0007742.  A portion of this research used resources at the High Flux Isotope Reactor, a DOE Office of Science User Facility operated by Oak Ridge National Laboratory.  The authors are thankful to Dr. Anil Sachdev at General Motors (GM) and Andrew Bobel for valuable discussions.

%\bibliography{references}

\end{document}